\documentclass[aps,prb,twocolumn,showpacs,groupedaddress,amsmath,amsymb,floatfix]{revtex4}

\usepackage{graphics,graphicx,epsfig}
\usepackage{dcolumn}
\usepackage{bm}

\begin{document}

\title{High-accuracy first-principles determination of the structural, vibrational
 and thermodynamical properties of diamond, graphite,  and derivatives}

\author{Nicolas Mounet} \email{nicolas.mounet@polytechnique.org}

\author{Nicola Marzari} \homepage{http://nnn.mit.edu/}
\affiliation{Department of Materials Science
and Engineering, Massachusetts Institute of Technology, Cambridge, MA,
USA}

\date{\today}

\begin{abstract}

The structural, dynamical, and thermodynamical properties of diamond, graphite and 
layered derivatives (graphene, rhombohedral graphite) are computed using a combination of
density-functional theory (DFT) total-energy calculations
and density-functional perturbation theory (DFPT) lattice dynamics at the GGA-PBE level.
Overall, very good agreement is found for the structural properties and
phonon dispersions, with the exception of the $c/a$ ratio in graphite and the associated
elastic constants and phonon dispersions. 
Both the C$_{33}$ elastic constant
and the $\Gamma$ to $A$ phonon dispersions 
are brought to close agreement with available data once the experimental $c/a$ is 
chosen for the calculations. 
The thermal expansion, the temperature dependence of the elastic moduli and the 
specific heat have been calculated via the quasi-harmonic 
approximation. Graphite shows a distinctive in-plane negative thermal-expansion 
coefficient that reaches the minimum
around room temperature, in very good agreement with experiments. Thermal contraction
in graphene is found to be three times as large; in both cases, ZA
acoustic modes are shown to be responsible for the contraction, in a direct manifestation of the 
membrane effect predicted by Lifshitz over fifty years ago.
\end{abstract}


\maketitle

\section{\label{sec1}Introduction}

The extraordinary variety of carbon allotropes, as well as their present and
potential applications in such diverse fields as nanoelectronics\cite{Novoselov} 
or bioengineering\cite{biosensors},
gives them a special place among all the elements.
Even excluding fullerenes, nanotubes, and their derivatives,
single crystalline diamond, graphite and
 graphene (i.e. a single graphite layer) still lack a complete
characterization of their thermodynamic stability under a broad range of conditions
(see e.g. Refs.~\onlinecite{Galli-Martin1,Galli-Martin2,Car-Nature,Graphite,handbook}
 and citations therein).
In this respect, vibrational
properties play a crucial role in determining the thermodynamic properties of the  bulk.
Indeed, diamond being a wide band gap
material (E$_g$= 5.5 eV), electronic excitations do not account
for thermal properties up to high
temperatures. Graphite and graphene are
semi-metals, but the gap vanishes only at the K point 
 where the two massless bands cross (see e.g. Ref.~\onlinecite{Dresselhaus}); thus,
electronic excitations can also be neglected in these materials,
and the phonon dispersions provide all the information that is needed to calculate 
thermodynamical quantities
such as the thermal expansion or specific heat.

The aim of this paper is to
provide a converged, accurate determination of the structural, dynamical,
and thermodynamical properties of diamond, graphite, graphene and rhombohedral
graphite from first-principles calculations.  Although the phonon spectrum of diamond 
and its thermal properties 
have been studied extensively with experiments
\cite{Warren-Yarnell,Slack-Bartram} and calculations\cite{Pavone-Karch},
the phonon spectrum of graphite is still under active 
investigation\cite{Maultzsch-Reich,Wirtz-Rubio}, as well as
its thermal properties. Graphite in-plane thermal expansion has long been
recognized to be negative\cite{Bailey-Yates,Nelson-Riley}, and it has even been 
suggested\cite{handbook,Nelson-Riley} that this may be due to the internal stresses
related to the large expansion in the $c$ direction (Poisson effect). 
To resolve some of the open questions, and to provide a coherent picture for these
materials, we used extensive
ab-initio density-functional theory (DFT) and density-functional perturbation theory 
(DFPT)\cite{Baroni-Gironcoli,Giannozzi-Gironcoli} calculations.
DFT is a very efficient and accurate tool to obtain ground-state and linear-response
properties, especially when  paired with plane-wave basis sets, which easily allow to reach full convergence
with respect to basis size, and
ultrasoft pseudo-potentials\cite{ultrasoft}
for optimal performance and transferability.
We adopted the PBE-GGA\cite{PBE} exchange-correlation functional,
at variance with most of the early studies on diamond\cite{Pavone-Karch,Xie-Chen,Favot} 
and especially graphite\cite{Boettger,Schabel-Martins,Wirtz-Rubio,Dubay-Kresse,Pavone-Bauer,Ye-Liu}, which have 
been performed using the local density approximation (LDA). GGA calculations have
appeared mostly for the cases of diamond (GGA-PBE, Ref.~\onlinecite{Favot}) and graphene 
(GGA-PBE, Refs.~\onlinecite{Maultzsch-Reich,Wirtz-Rubio}), 
with some data for graphite appearing in Refs.~\onlinecite{Martin-Lee,Ngoepe,Wirtz-Rubio,Mauri} 
(GGA-PBE). DFPT\cite{Baroni-Gironcoli,Giannozzi-Gironcoli} is then used
to compute the phonon frequencies at any arbitrary
wave-vector, without having to resort to the use
of supercells. The vibrational free energy
is calculated in the quasi-harmonic approximation (QHA)\cite{Pavone-Karch,Pavone},
to predict finite-temperature lattice
properties such as thermal expansion and specific heat.

To the best  of our knowledge, this is the first study 
on the thermal properties of graphite or graphene from first-principles.
For the case of diamond and graphene, calculations are fully ab-initio and do not
require any experimental input. For the case of graphite and rhombohedral graphite we argue
that the use of the experimental $c/a$ greatly improves the agreement with
experimental data. This experimental input is required since DFT,
in its current state of development, 
yields poor predictions for the interlayer interactions, dominated by Van Der
Waals dispersion forces not well described by local or semi-local exchange correlation
functionals (see  Refs.~\onlinecite{Langreth} and~\onlinecite{Kohn-Meir} for details; the agreement
between LDA predictions and experimental results for the $c/a$ ratio is fortuitous).
It is found that the weak interlayer bonding has a small
influence on most of the properties studied and that forcing the experimental
$c/a$ corrects almost all the remaining ones. This allowed us to
obtain results for all the materials considered that are in very good agreement 
with the available experimental data.

The article is structured as follows. We give a 
brief summary of our approach and definitions and introduce DFPT 
and the QHA in Section~\ref{sec2}. Our ground-state, zero-temperature
results for diamond, graphite, graphene and rhombohedral graphite are presented
in Section~\ref{sec3}:
Lattice parameters and elastic constants
from the equations of state in subsection~\ref{sec3:sub1}, phonon 
frequencies and vibrational density of states 
in subsection~\ref{sec3:sub2}, and first-principles, linear-response interatomic 
force constants 
in subsection~\ref{sec3:sub3}. The lattice thermal properties, such as
thermal expansion, mode Gr\"uneisen parameters,
and specific heat as obtained from the vibrational free energy are presented
in section~\ref{sec4}. Section~\ref{sec5} contains our final remarks.

\section{\label{sec2}Theoretical framework}

\subsection{\label{sec2:sub1}Basics of Density-Functional Perturbation Theory}

In density-functional theory\cite{Kohn-Sham,Hohenberg-Kohn} the ground state 
electronic density and wavefunctions of a crystal are found by
solving self-consistently a set of one-electron equations. In atomic
units (used throughout the article), these are

\begin{subequations}
\begin{equation}
(-\frac{1}{2}\mathbf{\nabla}^2 + V_{SCF}(\textbf{r})) | \psi_i \rangle
=\varepsilon_i |\psi_i \rangle, \label{KS1}
\end{equation}
\begin{equation}
V_{SCF}(\textbf{r})=\int \frac {n(\textbf{r}')}
{|\textbf{r}-\textbf{r}'|} d^3\textbf{r}' + \frac {\delta E_{xc}}
{\delta (n(\textbf{r}))} + V_{ion}(\textbf{r}), \label{KS2}
\end{equation}
\begin{equation}
n(\textbf{r})=\sum_i |\psi_i(\textbf{r})|^2
f(\varepsilon_F-\varepsilon_i), \label{KS3}
\end{equation}
\end{subequations}
where $f(\varepsilon_F-\varepsilon_i)$ is the occupation function,
$\varepsilon_F$ the Fermi energy, $E_{xc}$ the exchange-correlation
functional (approximated by GGA-PBE in our case), $n(\textbf{r})$ the
electronic-density, and $V_{ion}(\textbf{r})$ the ionic core potential
(actually a sum over an array of pseudo-potentials).
 
Once the unperturbed ground state is determined, phonon
frequencies can be obtained from the interatomic force
constants, i.e. the second derivatives at equilibrium of the total
crystal energy versus displacements of the ions:

\begin{eqnarray}
\nonumber &C_{\alpha i , \, \beta j} ( \textbf{R} - \textbf{R}' ) &= \left .
{\frac {\partial^2 E} {\partial u_{\alpha i}(\textbf{R}) \partial
u_{\beta j}(\textbf{R}') }} \right |_{equil} \\ \nonumber\\
\nonumber &&=C_{\alpha i , \, \beta
j}^{ion} ( \textbf{R} - \textbf{R}' )+C_{\alpha i , \, \beta j}^{elec}
( \textbf{R} - \textbf{R}' )\\ &&
\end{eqnarray}
Here \textbf{R} (\textbf{R}') is a Bravais lattice vector, i (j) indicates the
i$^{th}$ (j$^{th}$) atom of the unit cell, and $\alpha (\beta)$
represents the cartesian components. $C_{\alpha i , \, \beta j}^{ion}$
are the second derivatives\cite{Giannozzi-Gironcoli} of
Ewald sums corresponding to the ion-ion repulsion potential, while the 
electronic contributions $C_{\alpha i , \, \beta j}^{elec}$ are the second 
derivatives of the electron-electron and electron-ion terms in the ground
state energy. From the Hellmann-Feynman\cite{Giannozzi-Gironcoli} theorem one 
obtains:

\begin{eqnarray}
\nonumber &C_{\alpha i , \, \beta j}^{elec} ( \textbf{R} - \textbf{R}' ) =& \int
\left [ \frac {\partial n(\textbf{r}) } {\partial u_{\alpha i
}(\textbf{R})} \frac {\partial V_{ion}(\textbf{r})} {\partial u_{\beta
j}(\textbf{R}') } \right . \\ 
\nonumber && + \left . n_0(\textbf{r}) \frac {\partial^2
V_{ion}(\textbf{r})} {\partial u_{\alpha i} (\textbf{R}) \partial
u_{\beta j}(\textbf{R}') } \right ] d^3 \textbf{r} \\ &&
\end{eqnarray}
(where the dependence of both $n(\textbf{r})$ and
$V_{ion}(\textbf{r})$ on the displacements has been omitted for
clarity, and $V_{ion}(\textbf{r})$ is considered local).
 
It is seen that the electronic contribution can be
obtained from the knowledge of the linear response of the system to a displacement. 
The key assumption is then the Born-Oppenheimer
approximation which views a lattice vibration as a static
perturbation on the electrons. This is equivalent to say that the
response time of the electrons is much shorter than that of ions, that is,
each time ions are slightly displaced by a phonon, electrons
instantaneously rearrange themselves in the state of minimum energy of
the new ionic configuration. Therefore, static linear response
theory can be applied to describe the behavior of electrons upon a
vibrational excitation.

For phonon calculations, we consider a
periodic perturbation $\Delta V_{ion}$ of wave-vector \textbf{q},
which modifies the self-consistent potential $V_{SCF}$ by an amount 
$\Delta V_{SCF}$. The linear response in the charge density 
$\Delta n(\textbf{r})$ can be found using first-order
perturbation theory. If we consider its Fourier transform $\Delta
n(\textbf{q}+\textbf{G})$, and calling $\psi_{o,\textbf{k}}$
the one-particle wavefunction of an electron in the occupied band
``$o$'' at the point \textbf{k} of the Brillouin zone (and
$\varepsilon_{o,\textbf{k}}$ the corresponding eigenvalue), one can
get a self-consistent set of linear equations similar to
Eqs.~(\ref{KS1},\ref{KS2},\ref{KS3})\cite{Baroni-Giannozzi}:

\begin{subequations}
\begin{equation}
(\varepsilon_{o,\textbf{k}}+\frac{1}{2}\mathbf{\nabla}^2 -
V_{SCF}(\textbf{r})) \Delta \psi_{o,\textbf{k}+\textbf{q}} =
\hat{P}_e^{\textbf{k}+\textbf{q}} \Delta V_{SCF}^{\textbf{q}}
\psi_{o,\textbf{k}}
\end{equation}
\begin{equation}
\Delta n (\textbf{q}+\textbf{G}) = \frac {4} {V}
\sum_{\textbf{k},o} \langle \psi_{e,\textbf{k}} |
e^{-i(\textbf{q}+\textbf{G}) \cdot \textbf{r}}
\hat{P}_e^{\textbf{k}+\textbf{q}} | \Delta
\psi_{o,\textbf{k}+\textbf{q}} \rangle
\end{equation}
\begin{eqnarray}
\nonumber &\Delta V_{SCF}(\textbf{r})=& \int \frac {\Delta n(\textbf{r}')}
{|\textbf{r}-\textbf{r}'|} d^3\textbf{r}' + \Delta n(\textbf{r}) \left
[ \frac {d} {dn} \left ( \frac {\delta E_{xc}} {\delta (n(\textbf{r}))} 
\right ) \right ]_{n_0(\textbf{r})}\\\nonumber\\&& + \Delta V_{ion}(\textbf{r})
\end{eqnarray}
\end{subequations}
$\hat{P}_e^{\textbf{k}+\textbf{q}}$ refers to the projector on the
empty-state manifold at $\textbf{k}+\textbf{q}$, $V$ to the total
crystal volume, and \textbf{G} to any reciprocal lattice vector.  Note
that the linear response contains only Fourier components of wave
vector $\textbf{q}+\textbf{G}$, so we added a superscript \textbf{q} to
$\Delta V_{SCF}^{\textbf{q}}$. We have implicitly assumed for
simplicity that the crystal has a band gap and that pseudo-potentials
are local, but the generalization to metals\cite{Gironcoli} and to non-local
pseudo-potentials\cite{Giannozzi-Gironcoli} are all well established (see
Ref.~\onlinecite{Baroni-Gironcoli} for a detailed and complete review
of DFPT).

Linear-response theory allows us to calculate the response to any periodic 
perturbation; i.e. it allows direct access to the dynamical matrix 
related to the interatomic force constants via a Fourier transform:

\begin{equation}
\tilde{D}_{\alpha i, \, \beta j} (\textbf{q}) = \frac {1} {\sqrt {M_i
M_j}} \sum_{\textbf{R}} C_{\alpha i , \, \beta j} ( \textbf{R} ) \,
e^{-i \textbf{q}\cdot \textbf{R}}
\end{equation}
(where $M_i$ is the mass of the i$^{th}$ atom).

Phonon frequencies at any \textbf{q} are the solutions of the
eigenvalue problem:

\begin{equation}
\omega^2(\textbf{q}) u_{\alpha i}(\textbf{q}) = \sum_{\beta j}
u_{\beta j}(\textbf{q}) \tilde{D}_{\alpha i, \, \beta j} (\textbf{q})
\end{equation}
In practice, one calculates the dynamical matrix on a relatively
coarse grid in the Brillouin zone (say, a $8\times 8\times 8$
grid for diamond), and obtains the corresponding interatomic force
constants by inverse Fourier transform (in this example it would
correspond to a $8\times 8\times 8$ supercell in real space). Finally,
the dynamical matrix (and phonon frequencies) at any \textbf{q} point can
be obtained by Fourier interpolation of the real-space interatomic 
force constants.

\subsection{\label{sec2:sub2}Thermodynamical properties}

When no external pressure is applied to a crystal, the equilibrium
structure at any temperature T can be found by
minimizing the Helmholtz free energy $F(\{a_i\},T)=U-TS$ with respect to all its 
geometrical degrees of freedom $\{a_i\}$. If now the crystal is supposed to
be perfectly harmonic, F is the sum of the ground state total energy
and the vibrational free energy coming from the partition function (in
the canonical ensemble) of a collection of independent harmonic
oscillators. In a straightforward manner, it can be shown\cite{Maradudin} that:

\begin{eqnarray}
\nonumber & F(\{a_i\},T)& =E(\{a_i\})+ F_{vib}(T)\\ \nonumber \\ \nonumber &
&=E(\{a_i\})+\sum_{\textbf{q},j} \frac {\hbar \omega_{\textbf{q},j} } {2}\\
&& \quad + \, k_B T \sum_{\textbf{q},j} \ln \left ( 1-\exp \left ( -\frac 
{\hbar \omega_{\textbf{q},j}} {k_B T} \right ) \right ) \label{free}
\end{eqnarray}
where $E(\{a_i\})$ is the ground state energy and the sums run over
all the Brillouin zone wave-vectors and the band index $j$ of the phonon 
dispersion. The second term in the right hand side of Eq.(\ref{free})
is the zero-point motion.

If anharmonic effects are neglected, the
phonon frequencies do not depend on lattice parameters, therefore the
free energy dependence on structure is entirely contained in the
ground state equation of state $E(\{a_i\})$. Consequently the structure
 does not depend on temperature in a harmonic crystal.

Thermal expansion is recovered by introducing in Eq.(\ref{free}) the
 dependence of the phonon frequencies on the structural parameters $\{a_i\}$;
 direct minimization of the free energy
\begin{eqnarray}
\nonumber & F(\{a_i\},T)& =E(\{a_i\})+ F_{vib}(\omega_{\textbf{q},j}(\{a_i\}),T)\\
 \nonumber \\ \nonumber &
&=E(\{a_i\})+\sum_{\textbf{q},j} \frac {\hbar \omega_{\textbf{q},j}(\{a_i\})}{2}\\
\nonumber && \quad + \, k_B T \sum_{\textbf{q},j} \ln \left ( 1-\exp \left (
-\frac {\hbar \omega_{\textbf{q},j}(\{a_i\})} {k_B T} \right ) \right )\\ 
\label{free_quasi}
\end{eqnarray}
provides 
the equilibrium structure at any temperature T. This approach goes under the 
name quasi-harmonic approximation (QHA) and has been applied successfully to many bulk 
systems\cite{Pavone-Karch,Liu-metals,copper}. The linear thermal
expansion coefficients of the cell dimensions of a lattice are then

\begin{equation}
\alpha_i=\frac {1} {a_i} \frac {\partial a_i} {\partial T}
\end{equation}
The Gr\"uneisen formalism\cite{Barron-Collins} assumes a linear 
dependence of the phonon frequencies on the three orthogonal cell 
dimensions $\{a_i\}$; developing the ground state energy up to
second order, (thanks to the equation of state at $T=0 K$), one can get from the 
condition $ \left ( \frac {\partial F} {\partial a_i} \right )_T = 0$ the
alternative expression

\begin{equation}
\alpha_i=\sum_{\textbf{q},j} c_v(\textbf{q},j) \sum_k \frac { S_{i k}}
{V_0} \left ( \frac {-a_{0,k}} {\omega_{0,\textbf{q},j}} \left . \frac
{\partial \omega_{\textbf{q},j} } {\partial a_k} \right|_0 \right )
\end{equation}
We follow here the formalism of Ref.~\onlinecite{Schelling-Keblinski}:
 $c_v(\textbf{q},j)$ is the contribution to the specific heat from the mode 
$(\textbf{q},j)$, $S_{i k}$ is the elastic compliance matrix, and the 
subscript ``$0$'' indicates a quantity taken at the ground state lattice 
parameter.  The Gr\"uneisen parameter of the mode $(\textbf{q},j)$ is by
definition

\begin{equation}
\gamma_{k}(\textbf{q},j)=\frac {-a_{0,k}} {\omega_{0,\textbf{q},j}}
\left . \frac {\partial \omega_{\textbf{q},j} } {\partial a_k}
\right|_0
\end{equation}
For a structure which depends only on one lattice parameter $a$ (e.g.  diamond
or graphene) one then gets for the linear thermal expansion coefficient
\begin{equation}
\alpha=\frac {1} {d^2 B_0 V_0} \sum_{\textbf{q},j} c_v(\textbf{q},j)
\frac {-a_0} {\omega_{0,\textbf{q},j}} \left .  \frac {\partial
\omega_{\textbf{q},j} } {\partial a} \right|_0 \label{grun_1d}
\end{equation}
where $B_0$ is defined by $B_0=V_0 \frac {\partial^2 E} {\partial
V^2}$ ($V$ represents the volume of a three-dimensional
crystal such as diamond or the surface of a two-dimensional one like
graphene), d is the number of dimensions ($d=3$ for diamond, $d=2$ for
graphene), and $V_0$ is the volume (or the
surface) at equilibrium.

In the case of graphite there are two
lattice parameters: $a$ in the basal plane and $c$ perpendicular to the
basal plane, so that one gets

\begin{subequations}
\begin{eqnarray}
\nonumber &\alpha_a=& \frac {1} {V_0} \sum_{\textbf{q},j} c_v(\textbf{q},j)
\left ( (S_{11}+S_{12}) \frac {-a_0} {2 \omega_{0,\textbf{q},j}} \left . \frac
{\partial \omega_{\textbf{q},j} } {\partial a} \right|_0 \right . \\\nonumber\\
&&+ \left . S_{13} \frac {-c_0} {\omega_{0,\textbf{q},j}} \left . \frac {\partial
\omega_{\textbf{q},j} } {\partial c} \right|_0 \right )
\end{eqnarray}
\begin{eqnarray}
\nonumber &\alpha_c=& \frac {1} {V_0} \sum_{\textbf{q},j} c_v(\textbf{q},j)
\left ( S_{13} \frac {-a_0} {\omega_{0,\textbf{q},j}} \left .  \frac
{\partial \omega_{\textbf{q},j} } {\partial a} \right|_0 \right . \\\nonumber\\
&&+ \left . S_{33}\frac {-c_0} {\omega_{0,\textbf{q},j}} \left . \frac {\partial
\omega_{\textbf{q},j} } {\partial c} \right|_0 \right )
\end{eqnarray}
\end{subequations}
The mode Gr\"uneisen parameters provide useful insight to the thermal
expansion mechanisms. They are usually positive, since phonon frequencies decrease when
the solid expands, although some negative mode Gr\"uneisen
parameters for low-frequency acoustic modes can arise and sometimes compete with the
positive ones, giving a negative thermal expansion at low
temperatures, when only the lowest acoustic modes can be excited.

Finally, the heat capacity of the unit cell at constant volume can be obtained from 
$C_v=-T \left ( \frac {\partial^2 F_{vib}} {\partial T^2} \right )_V$ \cite{Maradudin}:

\begin{equation}
C_v=\sum_{\textbf{q},j} c_v(\textbf{q},j)=k_B \sum_{\textbf{q},j}
\left (\frac {\hbar \omega_{\textbf{q},j}} {2 k_B T} \right )^2 \frac
{1} {\sinh^2{\left (\frac {\hbar \omega_{\textbf{q},j}} {2 k_B T}
\right )} } \label{heat}
\end{equation}

\subsection{\label{sec2:sub3}Computational details}

All the calculations that follow were performed using the ESPRESSO\cite{PWSCF}
package, which is a full ab-initio DFT and DFPT code available under
the GNU Public License\cite{GNU}. We used a plane-wave basis set,
ultrasoft pseudo-potentials\cite{ultrasoft} from the 
standard distribution\cite{pseudopotential} (generated using a 
modified RRKJ\cite{RRKJ} approach), and the generalized gradient
approximation (GGA) for the exchange-correlation functional in its
PBE parameterization\cite{PBE}. We also used 
the local density
approximation (LDA) in order to compare some results between the 
two functionals.  In this case the
parameterization used was the one proposed by Perdew and
Zunger\cite{PZ}.

For the semi-metallic
graphite and graphene cases, we used 0.03 Ryd of
cold smearing \cite{Marzari-Vanderbilt}. We carefully and extensively
checked the convergence in the energy differences between different configurations
 and the phonon frequencies with respect to the wavefunction cutoff, the dual 
(i.e. the ratio between charge density cutoff and wavefunction cutoff), the
k-point sampling of the Brillouin zone, and the interlayer vacuum spacing
for graphene. Energy differences were converged within 5 meV/atom or better,
 and phonon frequencies within $1-2 \,\textrm{cm}^{-1}$. 
In the case of graphite and graphene phonon frequencies were 
converged with respect to the k-point sampling after having set the smearing 
parameter at 0.03 Ryd. Besides, values of the smearing
between 0.02 Ryd and 0.04 Ryd did not change the frequencies by more than 
$1-2 \,\textrm{cm}^{-1}$.
 
In a solid, translational invariance guaranties that three phonon frequencies
at $\Gamma$ will go to zero. In our GGA-PBE DFPT formalism
this condition is exactly satisfied only in the limit of infinite k-point 
sampling and full convergence with the plane-wave cutoff. For the case of
graphene and graphite we found in particular that an exceedingly large cutoff 
(100 Ryd) and dual (28) would be needed to recover phonon dispersions 
(especially around $\Gamma$ and the $\Gamma-A$ branch) with the tolerances
mentioned; on the other hand, application of the acoustic sum rule (i.e.
forcing the translational symmetry on the interatomic force constants) allows 
us to recover these highly converged calculations above with a more reasonable 
cutoff and dual.

Finally, the cutoffs we used were 40 Ryd for the wavefunctions in all the carbon
materials presented, with duals of 8 for diamond and 12 for graphite
and graphene. We used a $8\times 8 \times8$ Monkhorst-Pack k-mesh for diamond, 
$16\times 16\times 8$ for graphite, $16\times 16\times 4$ for rhombohedral 
graphite and $16\times 16\times 1$ for graphene. All these meshes were
unshifted (i.e. they do include $\Gamma$). Dynamical matrices were initially
calculated on a $8\times 8 \times8$ q-points mesh for diamond,
$8\times 8 \times4$ for graphite, $8\times 8 \times2$ for rhombohedral
graphite and $16\times 16 \times1$ for graphene.

Finally, integrations over the Brillouin zone for the vibrational free energy or the
heat capacity were done using phonon frequencies that were Fourier interpolated on 
much finer meshes. The phonon frequencies were usually computed at 
several lattice parameters and the results interpolated to get their dependence 
on lattice constants.

A final remark is that we were careful to use the same 
parameters (cutoffs, k-points sampling, smearing, etc.) in the determination of
the ground state equation of state and that of the phonon frequencies,
since these two terms need to be added in the free energy expression.

\section{\label{sec3}Zero-temperature results}

\subsection{\label{sec3:sub1}Structural and elastic properties}

We performed ground state total-energy calculations on diamond,
graphite, and graphene over a broad range of lattice parameters.  The
potential energy surface can then be fitted by an appropriate equation of 
state. The minimum gives the ground state equilibrium lattice
parameter(s). The second derivatives at that minimum are related to the bulk
modulus or elastic constants.

For the case of diamond we chose the Birch 
equation of state\cite{Ziambaras-Schroder} (up to the fourth order) to fit 
the total energy vs. the lattice constant $a$:

\begin{eqnarray}
\nonumber & E(a)= & -E_0+ \frac {9}{8}B_0 V_0 \left [ \left ( \frac
{a_0}{a} \right )^2 -1 \right ]^2 + A \left [ \left ( \frac {a_0}{a}
\right )^2 -1 \right ]^3 \\ & & + B \left [ \left ( \frac {a_0}{a}
\right )^2 -1 \right ]^4 + \mathcal{O}\left ( \left ( \frac {a_0}{a} \right )^2
-1 \right )^5
\end{eqnarray}
where $B_0$ is the bulk modulus, $V_0$ the primitive cell volume
($V_0=\frac {a^3} {4}$ here) and $A$ and $B$ are fit parameters. 
The Murnaghan equation of state or even 
a polynomial would fit equally well the calculations around the minimum of the
curve.  A best fit of this equation on our data gives us both the equilibrium
lattice parameter and the bulk modulus; our results are
summarized in Table~\ref{tab1}. The agreement with the
experimental values is very good, even after the zero-point motion and thermal
expansion are added to our theoretical predictions 
(see Section~\ref{sec4}).

\begin{figure}
\includegraphics*[scale=0.65]{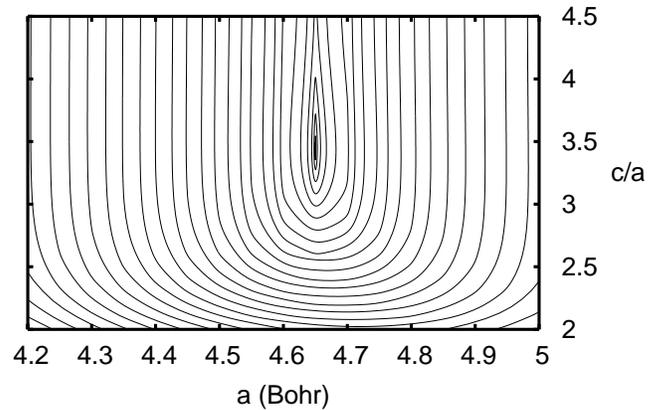}
\caption{\label{fig1}Contour plot of the ground state energy of graphite as a function of 
$a$ and $c/a$ (isoenergy contours are not equidistant).}
\end{figure}

\begin{figure}
\includegraphics*[scale=0.32]{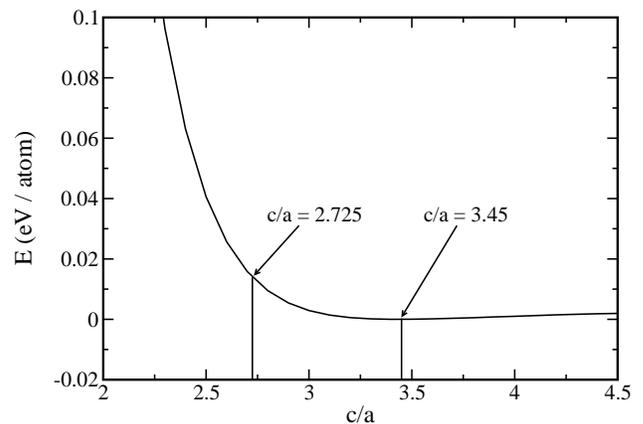}
\caption{\label{fig1bis}Ground state energy of graphite as a function of $c/a$ at 
fixed $a = 4.65 \, a.u. $. The theoretical (PBE) and the experimental $c/a$
are shown. The zero of energy has been set to the PBE minimum.}
\end{figure}

\begin{table}
\caption{\label{tab1} Equilibrium lattice parameter $a_0$ and bulk
modulus B$_0$ of diamond at the ground state (GS) and at 300 K (see 
Section~\ref{sec4}), compared to experimental values.}
\begin{ruledtabular}
\begin{tabular}{ccc}
                        &  Present calculation      & Experiment (300 K) \\ \hline
Lattice constant $a_0$  & 6.743 (GS) & 6.740
\footnote{Ref.~\onlinecite{Madelung}} \\
(a.u.)                  & 6.769 (300 K)        & \\ \hline
Bulk modulus $B_0$      & 432 (GS)    & 442 $\pm$ 2
\footnote{Ref.~\onlinecite{Grimsditch-Ramdas}}  \\
(GPa)                   & 422 (300 K)                   & 
\end{tabular}
\end{ruledtabular}
\end{table}
The ground state equation of state of graphene was fitted by a
$4^{th}$ order polynomial, and the minimum found for $a=4.654$ a.u.,
which is very close to the experimental in-plane lattice parameter of
graphite. The graphite equation of state was fitted by a
two-dimensional $4^{th}$ order polynomial of variables $a$ and $c$. To
illustrate the very small dependence of the ground state energy with
the $c/a$ ratio, we have plotted the results of our
calculations over a broad range of lattice constants in
Figs.~\ref{fig1} and~\ref{fig1bis}.


A few elastic constants can be obtained from the second derivatives of
this energy\cite{Boettger}:

\begin{subequations}\label{elas}
\begin{equation}
\textrm{Stiffness coefficients} \left\{ 
\begin{array} {crl}
&C_{11}+C_{12}=& \frac {1}{\sqrt{3} c_0} \frac {\partial^2 E}{\partial a^2}
\\ \\
&C_{33}=& \frac{2 c_0}{\sqrt{3} a_0^2} \frac {\partial^2 E} {\partial c^2}
\\ \\
&C_{13}=&\frac{1}{\sqrt{3} a_0} \frac {\partial^2 E} {\partial a \partial c}
\end{array} \right . \label{elas1}
\end{equation}
\begin{eqnarray} 
\nonumber \textrm{Tetragonal shear modulus} & C^t \,= 
&\frac {1} {6} \left [(C_{11}+C_{12}) \right . \\ \nonumber \\ 
&& +\left . 2C_{33}-4C_{13} \right ] \label{elas2} \\ \nonumber \\
\nonumber \textrm{Bulk modulus} & B_0 \,= 
&\frac {C_{33}(C_{11}+C_{12}) -2C_{13}^2} {6C^t} \\ \label{elas3}
\end{eqnarray}
\end{subequations}
\begingroup
\squeezetable
\begin{table*}
\caption{\label{tab2}Structural and elastic properties of graphite according to LDA,
GGA, and experiments}
\begin{ruledtabular}
\begin{tabular}{cccccc}
                              &  LDA fully  &  GGA fully  & GGA using           & GGA with             & Experiment \\
                              & theoretical & theoretical & exp. $c_0$      & 2$^{nd}$ derivatives & (300 K) \\
                              &             &             &in Eqs.~(\ref{elas1})&taken at exp. $c_0/a_0$& \\ \hline 
Lattice constant $a_0$(a.u.)  & 4.61        & 4.65        & 4.65                & 4.65(fixed)          & 4.65$\pm 0.003$\footnotemark[1]\\ 
$\frac {c_0}{a_0}$ ratio      & 2.74        & 3.45        & 3.45                & 2.725(fixed)         & 2.725$\pm 0.001$\footnotemark[1]\\
$C_{11}+C_{12}$ (GPa)         & 1283        & 976         & 1235                & 1230                 & 1240$\pm$40\footnotemark[2]\\
$C_{33}$ (GPa)                & 29          & 2.4         & 1.9                 & 45                   & 36.5$\pm$1 \footnotemark[2]\\ 
$C_{13}$ (GPa)                & -2.8        & -0.46       & -0.46               & -4.6                 & 15$\pm$5 \footnotemark[2]\\
$B_0$ (GPa)                   & 27.8        & 2.4         & 1.9                 & 41.2                 & 35.8 \footnotemark[3] \\ 
$C^t$ (GPa)                   & 225         & 164         & 207                 & 223                  & 208.8 \footnotemark[3]\\
\end{tabular}
\end{ruledtabular}
\footnotetext[1]{Refs.~\onlinecite{Zhao-Spain,Hanfland-Beister,Donohue},
as reported by Ref.~\onlinecite{Boettger}.}
\footnotetext[2]{Ref.~\onlinecite{Graphite}}
\footnotetext[3]{Ref.~\onlinecite{Blakslee-Proctor}, as reported by 
Ref.~\onlinecite{Boettger}}
\end{table*}
\endgroup
We summarize all our LDA and GGA results in Table~\ref{tab2}:
For LDA, both the lattice parameter $a_0$ and the $c_0/a_0$
ratio are very close to experimental data. Elastic constants were calculated 
fully from first-principles, in the sense that the second derivatives of the
energy were taken at the theoretical LDA $a_0$ and $c_0$, and that only
these theoretical values were used in Eqs.~(\ref{elas1}). Elastic constants are
 found in good agreement with experiments, except for the case of $C_{13}$ which comes
 out as negative (meaning that the Poisson's coefficient would be negative).

Fully theoretical GGA results (second column of Table~\ref{tab2}) compare 
poorly to experimental data except for the $a_0$ lattice constant, in very
good agreement with experiments. Using the experimental value for $c_0$ in
Eqs.~(\ref{elas}) improves only the value of $C_{11}+C_{12}$ (third column
of Table~\ref{tab2}).
Most of the remaining disagreement is related to the poor
value obtained for $c/a$; if the second derivatives in Eqs.~(\ref{elas1}) are taken at the 
experimental value for $c/a$ all elastic constants are accurately recovered except 
for $C_{13}$ (fourth column of Table~\ref{tab2}).

In both LDA and GGA, errors arise from the fact that 
Van Der Waals interactions between
graphitic layers are poorly described.
These issues can still be addressed within the framework of DFT
(as shown by Langreth and collaborators, Ref.~\onlinecite{Langreth}) at the 
cost of having a non-local exchange-correlation potential.

Zero-point motion and finite-temperature effects will be discussed in detail 
in Section~\ref{sec4}.

\subsection{\label{sec3:sub2}Phonon dispersion curves}

We have calculated the phonon dispersion relations for diamond,
graphite, rhombohedral graphite and graphene. For diamond and
graphene, we used the theoretical lattice parameter. For graphite,  we either 
used the theoretical $c/a$ or the experimental one ($c/a=2.725$). We will
comment extensively in the following on the role of $c/a$ on our calculated
properties.
 
\begin{figure*}
\includegraphics*[scale=0.65]{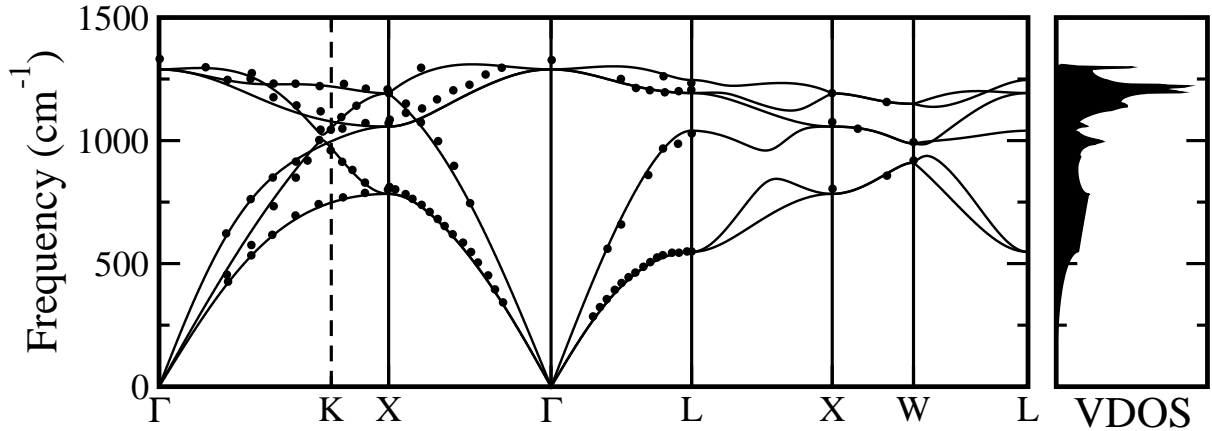}
\caption{\label{fig2}GGA ab-initio phonon dispersions 
(solid lines) and vibrational density of states (VDOS) for diamond.
Experimental neutron
scattering data from Ref.~\onlinecite{Warren-Yarnell} are shown for comparison (circles).}
\end{figure*}

\begin{figure*}
\includegraphics*[scale=0.6]{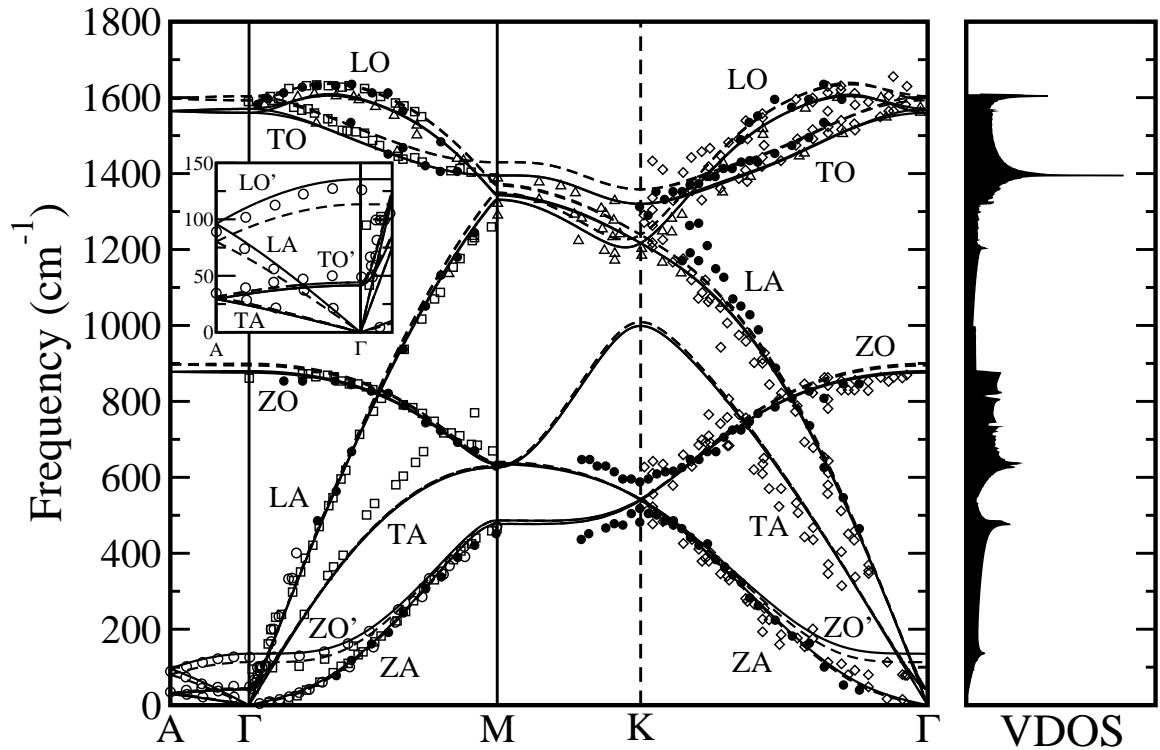}
\caption{\label{fig3}GGA (solid lines) and LDA (dashed line) ab-initio
 phonon dispersions for graphite, together with the GGA vibrational 
density of states (VDOS).
 The inset shows an enlargement of the low-frequency $\Gamma$-A region.
The experimental data are EELS (Electron Energy Loss
 Spectroscopy) from Refs.\onlinecite{Oshima-Aizawa},
 \onlinecite{Siebentritt-Pues}, \onlinecite{Yanagisawa-Tanaka}
 (respectively squares, diamonds, and filled circles), neutron
 scattering from Ref.~\onlinecite{Nicklow-Wakabayashi} (open
 circles), and x-ray scattering from
 Ref.~\onlinecite{Maultzsch-Reich} (triangles).  Data for
 Refs.~\onlinecite{Oshima-Aizawa} and \onlinecite{Yanagisawa-Tanaka}
 were taken from Ref.~\onlinecite{Wirtz-Rubio}.}
\end{figure*}

\begin{figure}
\includegraphics*[scale=0.35]{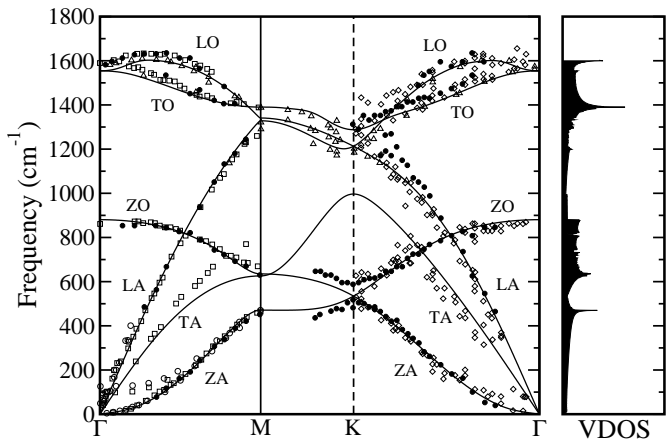}
\caption{\label{fig4}GGA ab-initio phonon dispersions for graphene
(solid lines). Experimental data for graphite are also shown, as in Fig.~\ref{fig3}.}
\end{figure}

\begin{figure}
\includegraphics*[scale=0.35]{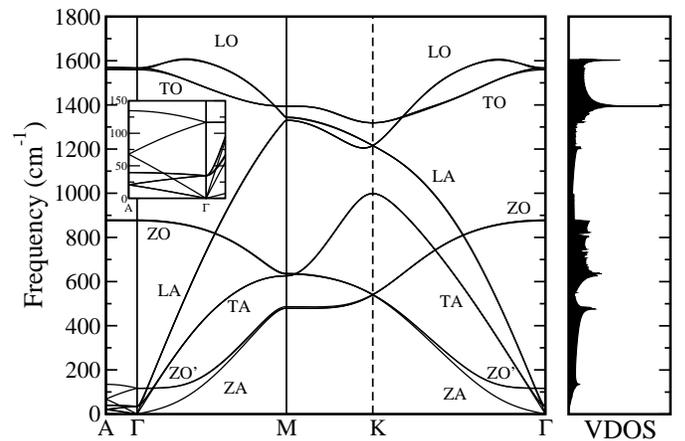}
\caption{\label{fig5}GGA ab-initio phonon dispersions for rhombohedral
graphite. The inset shows an enlargement of the low-frequency $\Gamma$-A region.}
\end{figure}

\begin{figure}
\includegraphics*[scale=0.35]{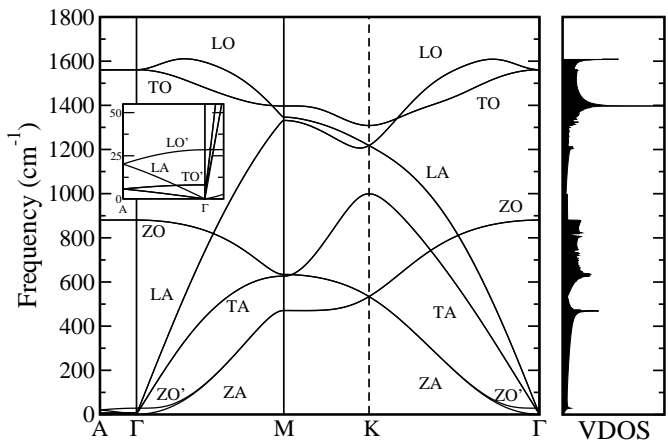}
\caption{\label{ph_GGA_theo}GGA ab-initio phonon dispersions for graphite 
at the theoretical $c/a$. 
The inset shows an enlargement of the low-frequency $\Gamma$-A region.}
\end{figure}

Finally we also calculated the phonon dispersions for rhombohedral graphite,
 which differs
from graphite only in the stacking of the parallel layers: in graphite the
stacking is ABABAB while it is ABCABC in rhombohedral graphite, and the latter 
unit cell contains six atoms instead of four.  We therefore used the same
in-plane lattice parameter and same interlayer distance as in graphite
(that is, a $\frac {c} {a}$ ratio multiplied by $\frac{3}{2}$
). Results are presented in Figs.~\ref{fig2}, \ref{fig3}, \ref{fig4}, 
\ref{fig5} and \ref{ph_GGA_theo}, together with the experimental data.

In Table~\ref{tab3} and \ref{tab4} we summarize our results at high-symmetry 
points and compare them with experimental data. In diamond, GGA produces 
softer modes than LDA~\cite{Pavone-Karch} on the whole (as expected),
particularly at $\Gamma$ (optical mode) and in the optical $\Gamma$-X branches.
For these, the agreement is somehow better in LDA; on the other hand the whole 
$\Gamma$-L dispersion is overestimated by LDA.

\begin{table}
\caption{\label{tab3} Phonon frequencies of diamond at the
 high-symmetry points $\Gamma$, X and L, in cm$^{-1}$.}
\begin{ruledtabular}
\begin{tabular}{ccccccccc}
                                         &$\Gamma_{O}$&$X_{TA}$&$X_{TO}$&$X_{LO}$&$L_{TA}$&$L_{LA}$&$L_{TO}$&$L_{LO}$
\\ \hline LDA\footnote{Ref.~\onlinecite{Pavone-Karch}}   & 1324
& 800    & 1094   & 1228   & 561   & 1080   & 1231   & 1275 \\
GGA\footnote{Present calculation}        & 1289        & 783    & 1057
& 1192  & 548    & 1040   & 1193   & 1246 \\
Exp.\footnote{Ref.~\onlinecite{Warren-Yarnell}}& 1332        & 807 &
1072   & 1184 & 550    & 1029   & 1206   & 1234 \\
\end{tabular}
\end{ruledtabular}
\end{table}
\begingroup \squeezetable
\begin{table*}
\caption{\label{tab4} Phonon frequencies of graphite and derivatives
 at the high-symmetry points A, $\Gamma$, M and K, in cm$^{-1}$. The
 lattice constants used in the calculations are also shown.}
\begin{ruledtabular}
\begin{tabular}{lcccccc}
& \multicolumn{3}{c}{Graphite} & Rhombo. graphite & Graphene & Graphite\\ \hline
Functional & LDA & GGA & GGA   & GGA              & GGA      & Experiment \\ \hline
In-plane lattice ct. $a_0$ & 4.61 a.u. & 4.65 a.u. & 4.65 a.u. & 4.65 a.u. & 4.65 a.u. & 4.65 a.u. \\ \hline 
Interlayer distance$/a_0$ & 1.36 & 1.725 & 1.36 & 1.36 & 15 & 1.36 \\ \hline
$A_{TA/TO'}$     & 31  & 6    & 29   &      &      & 35\footnotemark[1]\\ 
$A_{LA/LO'}$     & 80  & 20   & 96   &      &      & 89\footnotemark[1]\\
$A_{LO}$         & 897 & 880  & 878  &      &      & \\
$A_{TO}$         &1598 & 1561 & 1564 &      &      & \\ \hline
$\Gamma_{LO'}$   & 44  & 8    & 41   & 35   &      & 49\footnotemark[1]\\
$\Gamma_{ZO'}$   & 113 & 28   & 135  & 117  &      & 95\footnotemark[2], 126\footnotemark[1]\\
$\Gamma_{ZO}$    & 899 & 881  & 879  & 879  & 881  & 861\footnotemark[2]\\
$\Gamma_{LO/TO}$ &1593 & 1561 & 1559 & 1559 & 1554 & 1590\footnotemark[2], 1575\footnotemark[6] \\  
                 &1604 & 1561 & 1567 &  & \\ \hline
$M_{ZA}$         & 478 & 471  & 477  & 479  & 471  & 471\footnotemark[1], 465\footnotemark[2], 451\footnotemark[4]\\
$M_{TA}$         & 630 & 626  & 626  & 626  & 626  & 630\footnotemark[4]\\
$M_{ZO}$         & 637 & 634  & 634  & 635  & 635  & 670\footnotemark[2]\\
$M_{LA}$         & 1349& 1331 & 1330 & 1330 & 1328 & 1290\footnotemark[3]\\
$M_{LO}$         & 1368& 1346 & 1342 & 1344 & 1340 & 1321\footnotemark[3]\\
$M_{TO}$         & 1430& 1397 & 1394 & 1394 & 1390 & 1388\footnotemark[3], 1389\footnotemark[2]\\ \hline
$K_{ZA}$         & 540 & 534  & 540  & 535  & 535  & 482\footnotemark[4], 517\footnotemark[4], 530\footnotemark[5]\\ 
$K_{ZO}$         & 544 & 534  & 542  & 539  & 535  & 588\footnotemark[4], 627\footnotemark[5]\\ 
$K_{TA}$         & 1009& 999  & 998  & 998  & 997  & \\ 
$K_{LA/LO}$      & 1239& 1218 & 1216 & 1216 & 1213 & 1184\footnotemark[3], 1202\footnotemark[3]\\ 
$K_{TO}$& 1359& 1308 & 1319\footnotemark[7]& 1319& 1288\footnotemark[7] & 1313\footnotemark[4], 1291\footnotemark[5]\\
\end{tabular}
\end{ruledtabular}
\footnotetext[1]{Ref.~\onlinecite{Nicklow-Wakabayashi}}
\footnotetext[2]{Ref.~\onlinecite{Oshima-Aizawa}}
\footnotetext[3]{Ref.~\onlinecite{Maultzsch-Reich}}
\footnotetext[4]{Ref.~\onlinecite{Yanagisawa-Tanaka}}
\footnotetext[5]{Ref.~\onlinecite{Siebentritt-Pues}}
\footnotetext[6]{Ref.~\onlinecite{Tuinstra-Koenig}}
\footnotetext[7]{Note that a direct calculation of this mode with DFPT (instead of the Fourier interpolation result given here) leads to a significantly lower value in the case of graphite --- 1297 cm$^{-1}$ instead of 1319 cm$^{-1}$. This explains much of the discrepancy between the graphite and graphene result, since in the latter we used a denser q-points mesh. This effect is due to the Kohn anomaly occurring at K\cite{Mauri}.}
\end{table*}
\endgroup
The results on graphite require some comments. In Table~\ref{tab4} and Figs.~\ref{fig3},
 \ref{fig4}, \ref{fig5} and \ref{ph_GGA_theo}, modes are classified as follow: \
L stands for longitudinal polarization, T
for in-plane transversal polarization and Z for out-of-plane
transversal polarization. For graphite, a prime 
(as in LO') indicates an optical mode where the two atoms in each layer of 
the unit cell oscillate together and in phase opposition to the two atoms
 of the other layer. A non-primed optical mode is instead a mode where atoms 
inside the same layer are ``optical'' with respect to each other. 
Of course ``primed'' optical modes do not exist for graphene, since there is
only one layer (two atoms) per unit cell.

We observe that stacking has a negligible effect on all the frequencies
above 400 cm$^{-1}$, since both rhombohedral graphite and hexagonal graphite 
show nearly the same dispersions except for the $\Gamma$-A branch and the
in-plane dispersions near $\Gamma$. The in-plane part of the dispersions
is also very similar to that of graphene, except of course
for the low optical branches (below 400 cm$^{-1}$) that appear in graphite
and are not present in graphene.

For graphite as well as diamond GGA tends to make the high optical modes 
weaker while LDA makes them stronger than experimental values. 
The opposite happens for the low optical modes, and for the $\Gamma$-A 
branch of graphite; the acoustic modes show marginal differences 
and are in very good agreement with experiments. Overall, the agreement of both LDA and GGA 
calculations with experiments is very good and comparable to that between different measurements.

Some characteristic features of both diamond and graphite are well reproduced 
by our ab-initio results, such as the LO branch overbending and the 
associated shift of the highest frequencies away from $\Gamma$. Also, in the case of graphite, 
rhombohedral graphite and graphene, the quadratic dispersion of the in-plane ZA branch in the vicinity 
of $\Gamma$ is observed; this is a characteristic feature of the phonon dispersions of layered 
crystals\cite{Zabel,Lifshitz}, observed experimentally e.g. with neutron 
scattering\cite{Nicklow-Wakabayashi}.
Nevertheless, some discrepancies are found in graphite. The most obvious one is
along the $\Gamma$-M TA branch, where EELS\cite{Oshima-Aizawa} data show much 
higher frequencies than calculations. Additionally several EELS 
experiments\cite{Siebentritt-Pues,Yanagisawa-Tanaka} report a gap between
the ZA and ZO branches at K while these cross each other in all the 
calculations. In these cases the disagreement could come either from a
failure of DFT within the approximations used or from imperfections in the crystals
used in the experiments.

There are also discrepancies between experimental data,
in particular in graphite for the LA branch around K: EELS data from 
Ref.~\onlinecite{Siebentritt-Pues} agree with our ab-initio results while those from 
Ref.~\onlinecite{Yanagisawa-Tanaka} deviate from them.

Finally, we should stress again the dependence of the graphite phonon
 frequencies on the in-plane 
lattice parameter and $c/a$ ratio. The results we have analyzed so far were 
obtained using the theoretical in-plane lattice parameter $a$ and the 
experimental $c/a$ ratio for both GGA and LDA. Since the LDA theoretical
$c/a$ is very close to the experimental one (2.74 vs. 2.725) and the interlayer bonding 
is very weak, these differences do no matter. However this is not the case 
for GGA, as the theoretical $c/a$ ratio is very different from the experimental
one (3.45 vs. 2.725). Fig.~\ref{ph_GGA_theo} and the second column of 
Table~\ref{tab4} show results of GGA calculations performed at the
theoretical $c/a$. Low frequencies (below 150 cm$^{-1}$)
between $\Gamma$ and A are 
strongly underestimated, as are the ZO' modes between $\Gamma$ and M, while
the remaining branches are barely affected.

The high-frequency optical modes are instead strongly
 dependent on the in-plane lattice constant. The difference between the 
values of $a$ in LDA and GGA explains much
of the discrepancy between the LDA optical modes and
 the GGA ones. Indeed, a LDA calculation
performed at $a=4.65$ a.u. and $c/a=2.725$ (not shown here) brings the phonon
frequencies of these modes very close to the GGA ones obtained
with the same parameters,
while lower-energy modes (below 1000 cm$^{-1}$) are hardly affected.

Our final choice to use the theoretical in-plane 
lattice parameter and the experimental $c/a$ seems to strike a balance between the
need of theoretical consistency and that of accuracy. Therefore, the 
remaining of this section is based on calculations performed using the parameters 
discussed above ($a=4.61$ for LDA, $a=4.65$ for GGA and $c/a=2.725$ 
in each case).

Elastic constants can be extracted from the data on sound velocities. Indeed, the latters are the slopes 
of the dispersion curves in the vicinity  of $\Gamma$ and can be expressed 
as the square root of linear combinations of
elastic constants (depending on the branch considered) over the
density (see Ref.~\onlinecite{Kittel} for details).
We note in passing that we computed the density
consistently with the geometry used in the calculations (see Table~\ref{tab4} for
details, first column for LDA and third one for GGA), and not the 
experimental density. Our results are shown in Table~\ref{tab5}.

\begin{table}
\caption{\label{tab5} Elastic constants of diamond and graphite as calculated from 
the phonon dispersions, in GPa.}
\begin{ruledtabular}
\begin{tabular}{lccccc}
& \multicolumn{2}{c}{Diamond} & \multicolumn{3}{c}{Graphite} \\ \hline
Functional & GGA& Exp.              & LDA & GGA & Exp.\\ \hline 
$C_{11}$   &1060&1076.4 $\pm$ 0.2\footnotemark[2]&1118&1079&1060 $\pm$ 20\footnotemark[1]\\
$C_{12}$   & 125&125.2 $\pm$ 2.3\footnotemark[2] &235 & 217& 180 $\pm$ 20\footnotemark[1]\\ 
$C_{44}$   & 562&577.4 $\pm$ 1.4\footnotemark[2] &4.5 & 3.9& 4.5 $\pm$ 0.5\footnotemark[1]\\
$C_{33}$   & -  & -                              &29.5&42.2& 36.5$\pm$1 \footnotemark[1] 
\end{tabular}
\end{ruledtabular}
\footnotetext[1]{Ref.~\onlinecite{Graphite}}
\footnotetext[2]{Ref.~\onlinecite{Grimsditch-Ramdas}}
\end{table}
The overall agreement with experiment is good to very good. LDA leads to larger
 elastic constants, as expected from the general tendency to ``overbind'', 
but still agrees well with experiment. For diamond, the agreement is particularly
 good. As for $C_{13}$ in graphite, it is quite difficult to obtain it from the dispersion
 curves since it enters the sound velocities only in a linear combination
involving other elastic constants, for which the error is almost
comparable to the magnitude of $C_{13}$ itself.

An accurate description of the phonon dispersions allow us to predict the low-energy
structural excitations and thus several thermodynamical quantities. Before
exploring this in Section~\ref{sec4}, we want to discuss the nature and decay 
of the interatomic force constants in carbon based materials.

\subsection{\label{sec3:sub3}Interatomic force constants}

\begin{figure}
\includegraphics*[scale=0.35]{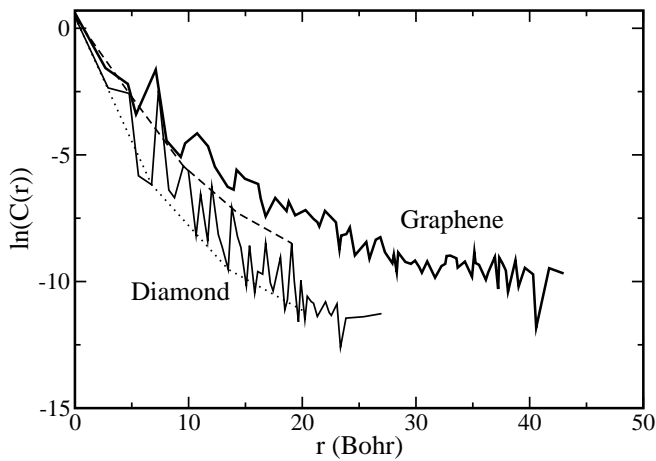}
\caption{\label{fig6}Decay of the norm of the interatomic force constants
as a function of distance for diamond (thin solid line) and graphene (thick
solid line), in a semi-logarithmic scale. The dotted and dashed lines show
the decay for diamond along the (100) and (110) directions.}
\end{figure}

As explained in Section~\ref{sec2:sub1}, the interatomic force constants 
$C_{i , \, j} ( \textbf{R} - \textbf{R}' )$ are obtained
in our calculations from the Fourier transform of the dynamical matrix 
$\tilde{D}_{i, \, j} (\textbf{q})$ calculated on a regular
mesh inside the Brillouin zone ($8\times 8 \times8$ for diamond, 
$8\times 8 \times4$ for graphite and $16\times
16 \times 1$ for graphene). This procedure is exactly equivalent (but much
more efficient) than calculating the interatomic force constants with frozen
phonons (up to 47 neighbors in diamond and 74 in graphene). 
At a given $\textbf{R}$, $C_{i , \, j} ( \textbf{R})$ is actually a $2^{nd}$ order
 tensor, and the decay of its norm (defined as the square root of the sum of 
the squares of all the matrix elements) with distance is a
good measure to know the effect of distant neighbors.  In Fig.~\ref{fig6}
 we have plotted the natural logarithm
of such a norm with respect to the distance from a given atom, for
diamond and graphene. The norm has been averaged on all the neighbors
located at the same distance before taking the logarithm.

The force-constants decay in graphene is slower than in
diamond, and it depends much less on direction. In diamond decay along
(110) is much slower than in other directions due to long-range elastic effects
along the covalent bonds. This long-range decay is also responsible for the 
flattening of the phonon dispersions in zincblende and diamond semiconductors
along the K-X line (see Fig.~\ref{fig2} and Ref.~\onlinecite{Giannozzi-Gironcoli},
for instance).

\begin{figure}
\includegraphics*[scale=0.35]{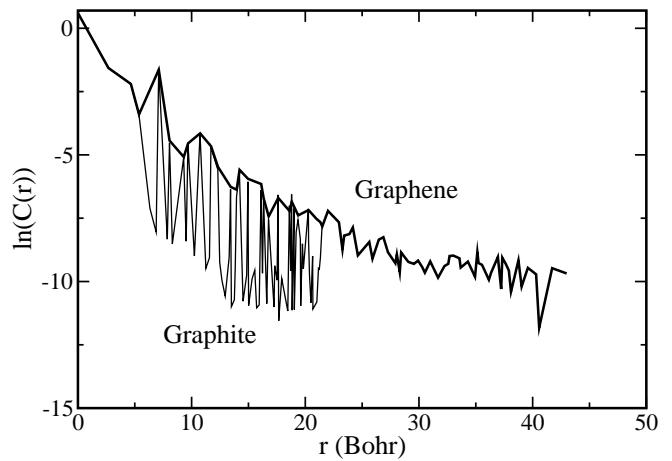}
\caption{\label{fig7}Decay of the norm of the interatomic force constants
as a function of distance for graphite (thin solid line) and graphene (thick
solid line).}
\end{figure}

In Fig.~\ref{fig7} we show
the decay plot for graphite and graphene, averaged over all 
directions. The graphite interatomic force constants include values corresponding 
to graphene (in-plane nearest neighbors) and smaller values corresponding to the weak 
interlayer interactions.

It is interesting to assess the effects of the truncation of these interatomic 
force constants on the phonon dispersion curves. This can be done by replacing the force
constants corresponding to distant neighbors by zero.
In this way the relevance of short-range and long-range contributions can be examined. 
The former are relevant for short-range force-constant 
models such as the VFF (Valence Force Field)\cite{Dresselhaus} or the 4NNFC 
(4$^{th}$ Nearest-Neighbor Force Constant)\cite{Aizawa-Souda} used e.g. in graphene.
Note however that a simple truncation is not comparable to the VFF or 
4NNFC models, where effective interatomic force constants would be renormalized.

\begin{figure}
\includegraphics*[scale=0.3]{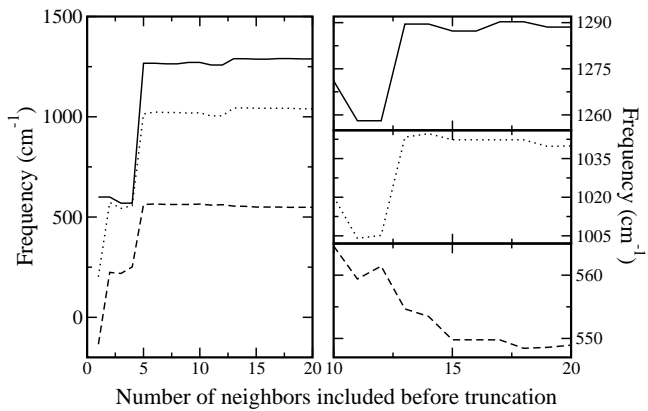}
\caption{\label{fig8}Phonon frequencies of diamond as a function of the number of
neighbors included in the interatomic force constants: 
$\Gamma_O$ (solid line), $X_{TO}$ (dotted line), and 
$L_{TA}$ (dashed line).}
\end{figure}

\begin{figure}
\includegraphics*[scale=0.3]{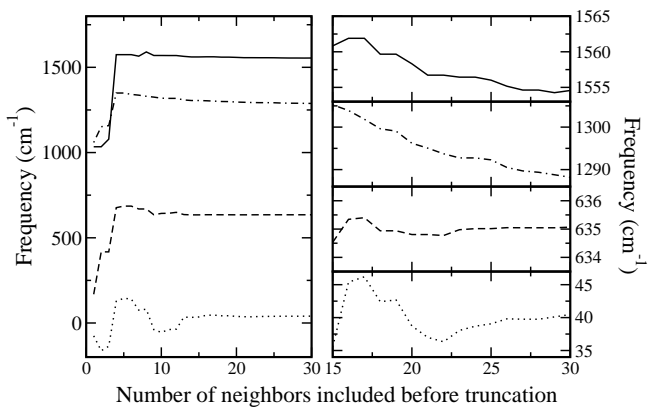}
\caption{\label{fig9}Phonon frequencies of graphene as a function of the number of
neighbors included in the interatomic force constants:
$\Gamma_{LO/TO}$ (solid line), 
$K_{TO}$ (dot-dashed), $M_{ZO}$ (dashed), and for the dotted line a
phonon mode in the ZA branch one-fourth along the $\Gamma$ to M line.}
\end{figure}

Figs.~\ref{fig8} and \ref{fig9} show the change in
frequency for selected modes in diamond and graphene as a function of the
truncation range. The modes we chose  are those most strongly
affected by the number of neighbors included.

For diamond, our whole
supercell contains up to 47 neighbors, and the graph shows
only the region up to 20 neighbors included, since the
selected modes do not vary by more than $1 \, \text{cm}^{-1}$
after that. With 5 neighbors, phonon frequencies are already near
their converged value, being off by at worst 4\% off;  
very good accuracy ($5 \, \text{cm}^{-1}$) is obtained with 13 neighbors.

For graphene, our $16\times 16 \times 1$ supercell contains up to 74 neighbors, but after 
the 30$^{th}$ no relevant changes occur. At least 4 neighbors are needed for the optical modes 
to be converged within 5-8\%. Some acoustic modes require more neighbors, 
as also pointed out in Ref.~\onlinecite{Dubay-Kresse}. As can be seen in Fig.~\ref{fig9}, the frequency of 
some ZA modes in the $\Gamma$-M branch (at about one fourth of the branch) oscillates
strongly with the number of neighbors included, and can even become imaginary when less than
13 are used, resulting into an instability of the crystal.
This behavior does not appear in diamond.
Also, the $K_{TO}$ mode keeps varying in going down from 20 to 30 neighbors, though this effect remains small
($8-9 \, \text{cm}^{-1}$). This drift could signal the presence of a 
Kohn anomaly\cite{Kohn}. Indeed, at the
 $K$ point of the Brillouin zone the electronic band gap vanishes 
in graphene, so that a singularity arises in the highest optical phonon mode. 
Therefore a finer q-point mesh is needed around this point, and longer-ranged 
interatomic force constants. This effect is discussed in detail in Ref.~\onlinecite{Mauri}.


\section{\label{sec4}Thermodynamical properties}

\begin{figure}
\includegraphics*[scale=0.35]{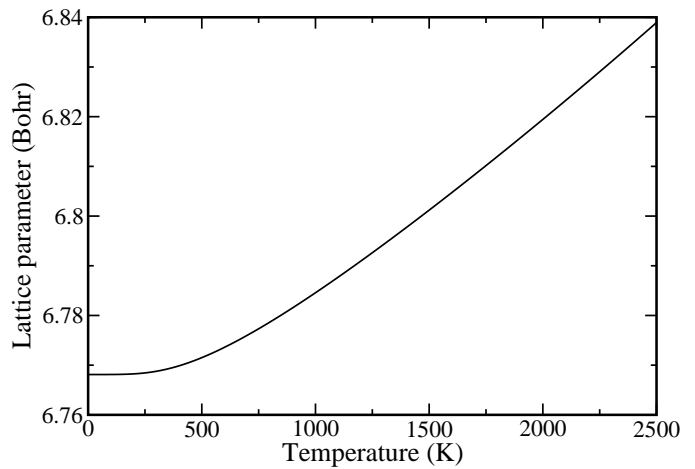}
\caption{\label{fig10}Lattice parameter of diamond
as a function of temperature}
\end{figure}

\begin{figure}
\includegraphics*[scale=0.35]{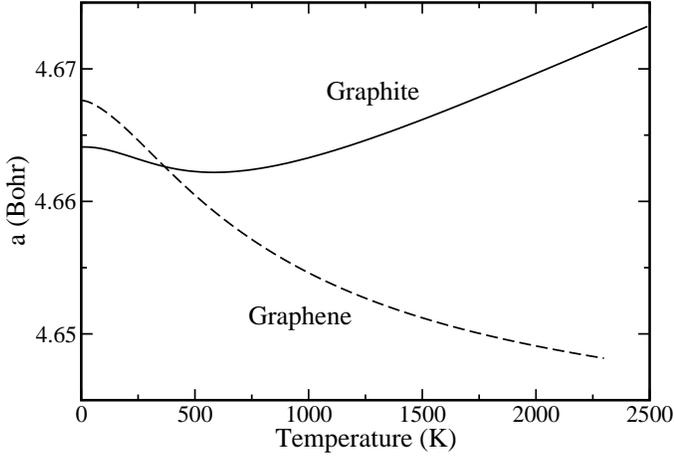}
\caption{\label{fig11}In-plane lattice parameter of graphite (solid
line) and graphene (dashed line)
as a function of temperature}
\end{figure}

\begin{figure}
\includegraphics*[scale=0.35]{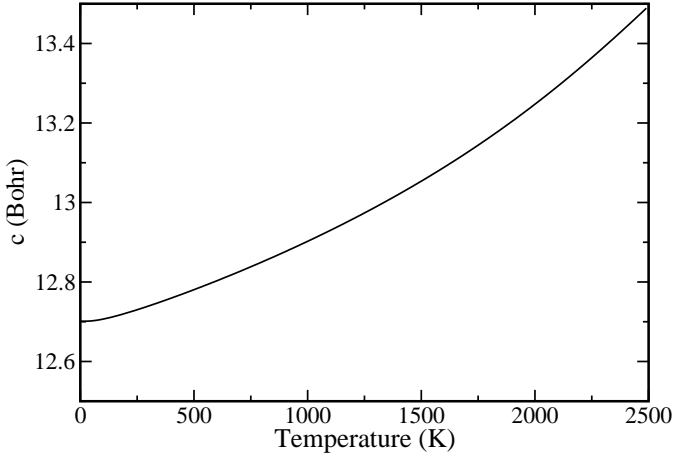}
\caption{\label{fig12}Out-of-plane lattice parameter of graphite as a function
of temperature}
\end{figure}

We present in this final section our results on the thermodynamical properties
of diamond, graphite and graphene using the quasi-harmonic
approximation and phonon dispersions at the GGA level. As outlined in
Section~\ref{sec2:sub2} we first perform a direct minimization over the 
lattice parameter(s) $\{a_i\}$ of the vibrational free energy $F(\{a_i\},T)$ 
(Eq.~\ref{free_quasi}).
 This gives us, for any temperature T, the equilibrium lattice parameter(s), shown 
in Figs.~\ref{fig10}, \ref{fig11} and \ref{fig12}. For
diamond and graphene, we used in Eq.~\ref{free_quasi} the equations of state 
obtained from the ground state calculations
presented in Section~\ref{sec3:sub1}. For graphite this choice would not be
useful or accurate, since the theoretical $c/a$ is much larger than the
experimental one. So we \textit{forced} the equation of state to be a minimum 
for $a$=4.65 a.u. and $\frac {c} {a}$=2.725 (fixing
only $c/a$ and relaxing $a$ would give $a$=4.66 a.u., with negligible effects
on the thermal expansion). In particular, our ``corrected'' equation of state
is obtained by fitting with a fourth order polynomial the true equation of 
state around the experimental $a$ and $c/a$, and then dropping from this 
polynomial the linear order terms. Since the second derivatives of the 
polynomial are unchanged, this is to say we keep the elastic constants unchanged. 
The only input from experiments remains the $c/a$ ratio. We have also checked the effect of imposing
to $C_{13}$ its experimental value ($C_{13}$ is the elastic constant that is less 
accurately predicted), but the
changes were small.

The dependence of the phonon frequencies on the lattice
parameters was determined by calculating the whole phonon dispersions
at several values and interpolating these in between. For
diamond and graphene we used four different values of $a$ (from 6.76 to 
6.85 a.u. for diamond, and from 4.654 to 4.668 a.u. for graphene) and interpolated 
them with a cubic polynomial. For graphite, since the
minimization space is two dimensional, we restricted ourselves to a linear
interpolation and calculated the phonon dispersions at three different 
combinations of the lattice constants : $(a,c/a)$=(4.659,2.725), (4.659,2.9)
and (4.667,2.725).

Before considering thermal expansion, we examine the zero-point motion. 
Indeed, lattice parameters at 0 K are different from their ground state values. 
The effects of the thermal expansion (or contraction) up to about 1000 K 
are small compared to the zero-point expansion of the lattice
parameters. In diamond, $a$ expands from 6.743 a.u. (ground state value)
to 6.768 a.u., a difference of 0.4\%.  For graphene, $a$ is
4.654 a.u. at the ground state and 4.668 a.u. with zero-point motion corrections
(+0.3\%); for graphite $a$ increases from 4.65 to 4.664 a.u. (+0.3\%)
and $c$ from 12.671 to 12.711 (+0.3\%). The increase is similar in
each case, and comparable to the discrepancy between experiments and GGA or LDA 
ground states.

\begin{figure}
\includegraphics*[scale=0.35]{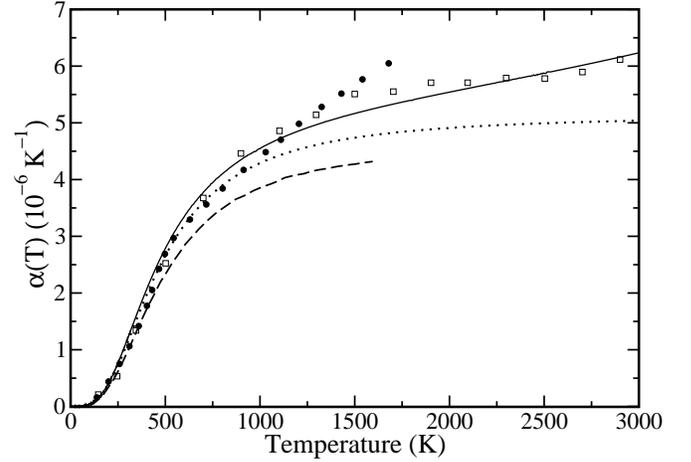}
\caption{\label{fig13}Coefficient of linear thermal expansion for diamond as a 
function of
temperature. We compare our QHA-GGA ab-initio calculations (solid line)
to experiments (Ref.~\onlinecite{Slack-Bartram}, filled circles),
a path integral Monte-Carlo study using a Tersoff
empirical potential (Ref.~\onlinecite{Herrero-Ramirez}, open squares) and 
the QHA-LDA study by Pavone \textit{et al}\cite{Pavone-Karch} (dashed line). 
The QHA-GGA thermal expansion calculated using the
Gr\"uneisen equation (Eq.~\ref{grun_1d}) 
is also shown (dotted line).}
\end{figure}

\begin{figure}
\includegraphics*[scale=0.35]{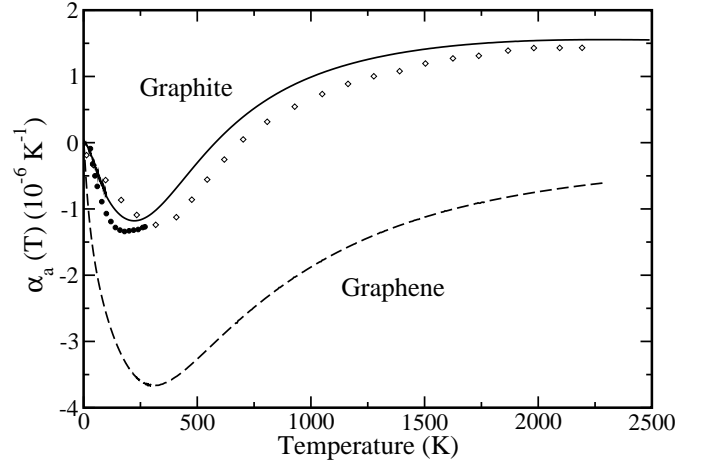}
\caption{\label{fig14}In-plane coefficient of linear thermal expansion 
as a function of temperature for graphite (solid line) and graphene (dashed line)
from our QHA-GGA ab-initio study.  The experimental results for graphite are from
Ref.~\onlinecite{Bailey-Yates} (filled circles) and
Ref.~\onlinecite{handbook} (open diamonds).}
\end{figure}

\begin{figure}
\includegraphics*[scale=0.35]{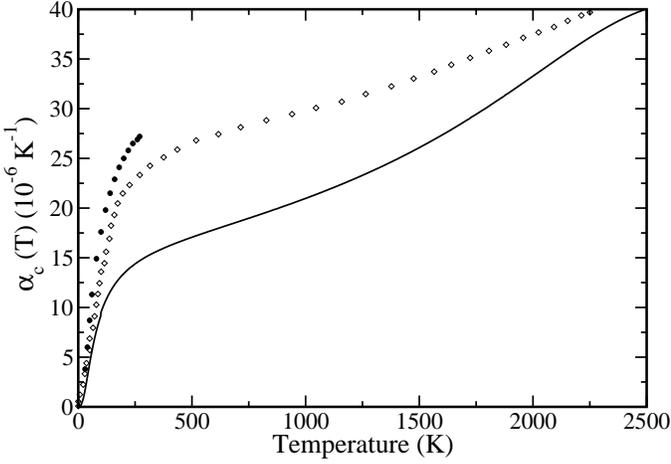}
\caption{\label{fig15}Out-of-plane coefficient of linear thermal expansion
as a function of temperature for graphite from our QHA-GGA ab-initio study (solid line).  
The experimental
results are from Ref.~\onlinecite{Bailey-Yates} (filled circles), and
Ref.~\onlinecite{handbook} (open diamonds).}
\end{figure}

The coefficients of linear thermal expansion
 at any T are obtained by numerical differentiation of the previous data. Results are
shown in Figs.~\ref{fig13}, \ref{fig14} and \ref{fig15}.

For the case of diamond, we have also plotted the linear thermal expansion
coefficient calculated using the Gr\"uneisen formalism (Eq.~\ref{grun_1d}) 
instead of directly minimizing the free energy. While at low temperature 
the two curves agree, a discrepancy becomes notable above 1000 K,
and direct minimization should be performed.
This difference between Gr\"uneisen theory and direct minimization
seems to explain much of the discrepancy between the calculations of 
Ref.~\onlinecite{Pavone-Karch} and our
results. Finally a Monte-Carlo path integral study by Herrero and
Ram\'irez \cite{Herrero-Ramirez}, which does not use the QHA,
gives very similar results.

For graphite, the in-plane coefficient of linear thermal expansion 
slightly overestimates the experimental values,
but overall the agreement remains excellent, even at high temperatures.
Out-of-plane, the agreement holds well up to 150 K, after which the
coefficient of linear thermal expansion is underestimated by about 30\% at 1000 K.

In-plane, the coefficient of linear thermal expansion is confirmed to
be negative from 0 to about 600 K. This feature, absent in
diamond, is much more apparent in graphene, where the coefficient of linear 
thermal expansion keeps being negative up to 2300 K. This thermal contraction
will likely appear also in single-walled nanotubes (one graphene sheet rolled on itself)
\cite{Mounet-Marzari}. Some molecular dynamics
calculations\cite{Schelling-Keblinski,Tomanek} have already pointed out this 
characteristic of SWNT.

\begin{figure}
\includegraphics*[scale=0.35]{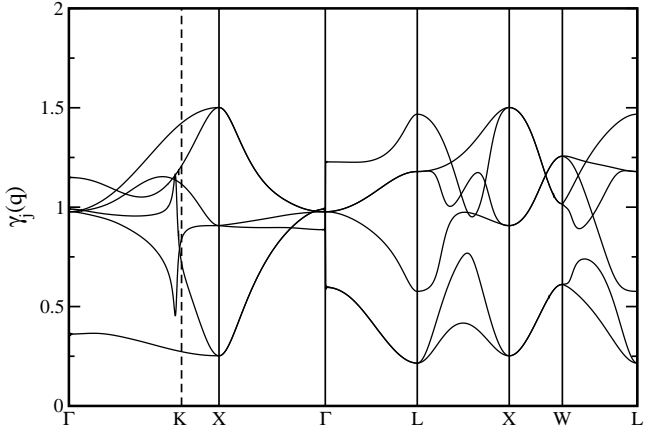}
\caption{\label{fig16}Ab-initio mode Gr\"uneisen parameters for diamond.}
\end{figure}

\begin{figure}
\includegraphics*[scale=0.35]{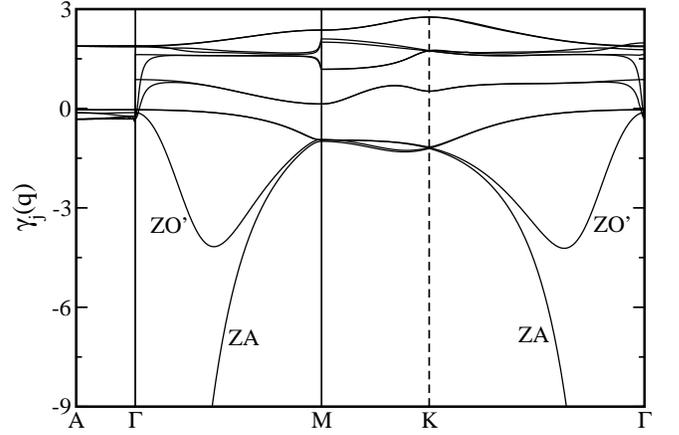}
\caption{\label{fig17}Ab-initio in-plane mode Gr\"uneisen parameters for graphite.}
\end{figure}

\begin{figure}
\includegraphics*[scale=0.35]{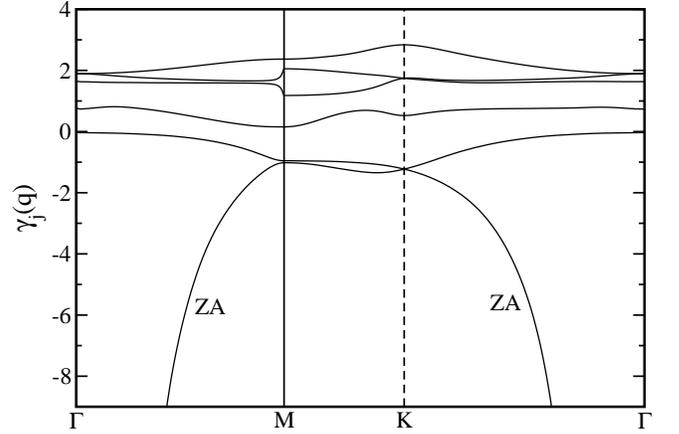}
\caption{\label{fig18}Ab-initio mode Gr\"uneisen parameters for graphene.}
\end{figure}

\begin{figure}
\includegraphics*[scale=0.35]{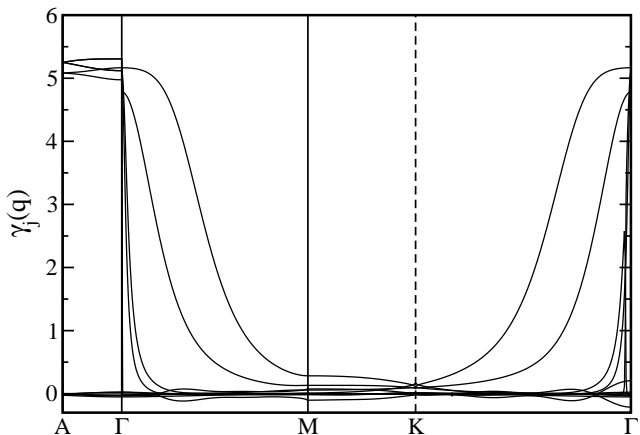}
\caption{\label{fig19}Ab-initio out-of-plane mode Gr\"uneisen parameters for graphite.}
\end{figure}

To further analyze thermal contraction, we plotted in
Figs. \ref{fig16}, \ref{fig17}, \ref{fig18} and \ref{fig19} the mode Gr\"uneisen
parameters (see Section \ref{sec2:sub2}) of diamond, graphene and graphite. 
These have been obtained 
from an interpolation of the phonon frequencies by a quadratic (or linear, 
for graphite) polynomial of the lattice constants, and computed at the ground
state lattice parameter.

The diamond Gr\"uneisen parameters have been already calculated with LDA 
(see Refs.~\onlinecite{Pavone-Karch,Xie-Chen}); our GGA results agree very well with
these. In particular, all the Gr\"uneisen parameters are shown to be
positive (at odds with other group IV semiconductors such as Si or Ge). 
The situation is very different in graphite and graphene, where some bands 
display large and negative Gr\"uneisen parameters (we have used the definition 
$\gamma_j (\textbf{q})=- \frac {a} {2 \omega_j(\textbf{q})} \frac {d
\omega_j(\textbf{q})} {d a}$).

While not visible in the figure, the Gr\"uneisen parameter for the lowest 
acoustic branch of graphite becomes as low as -40, and as low
as -80 for graphene. Therefore, at low temperatures (where most
optical modes with positive Gr\"uneisen parameters are still not
excited) the contribution from the negative Gr\"uneisen parameters will
be predominant and thermal expansion (from Eq.~\ref{grun_1d}) negative.
 
The negative Gr\"uneisen parameters correspond to the lowest
transversal acoustic (ZA) modes, and in the case of graphite to the (ZO') modes, 
which can be described as ``acoustic'' inside the layer and optical out-of-plane
(see Section~\ref{sec3:sub2}). Indeed, the phonon frequencies for such modes 
increase when the in-plane lattice parameter is increased, contrary to
the usual behavior, because the layer is more ``stretched'' when $a$ is
increased, and atoms in that layer will be less free to move in the z
direction (just like a rope that is stretched will have vibrations of
smaller amplitude, and higher frequency). In graphite these parameters are
less negative because of the interaction between layers: atoms are
less free to move in the z-direction than in the case of graphene.

This effect, known as the ``membrane effect'', was predicted by Lifshitz\cite{Lifshitz} in 1952, when
he pointed out the role of these ZA modes (also called ``bending modes'') in layered materials.
In particular, several recent studies have highlighted the relevance of these modes to the thermal
properties of layered crystals such as graphite, boron nitride and gallium 
sulfide\cite{Belenkii,Abdullaev,Abdullaev2}.

The knowledge of the equilibrium lattice constant(s) at any temperature allows
us also to calculate the dependence of elastic constants on temperature. To do so we 
calculated the second derivatives of the free energy (Eq.~\ref{free_quasi}) 
vs. lattice constant(s) at the finite-temperature equilibrium lattice parameter(s).
 We checked that this was equivalent to a best fit of the
free energy at T around the equilibrium lattice parameter(s).

\begin{figure}
\includegraphics*[scale=0.32]{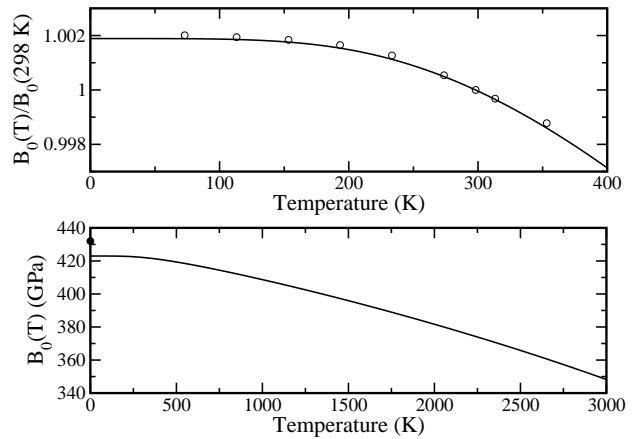}
\caption{\label{fig20}Lower panel: Bulk modulus $B_{0}(T)$ of diamond as a function of temperature.
The filled circle 
indicates the value of the bulk modulus (as in Table~\ref{tab1}) before accounting for zero-point motion.
Upper panel: 
theoretical (solid line) and experimental values (Ref.~\onlinecite{McSkimin}, open circles)
for the ratio between $B_{0}(T)$ and $B_{0}(298 K)$
in the low temperature regime.}
\end{figure}

\begin{figure}
\includegraphics*[scale=0.32]{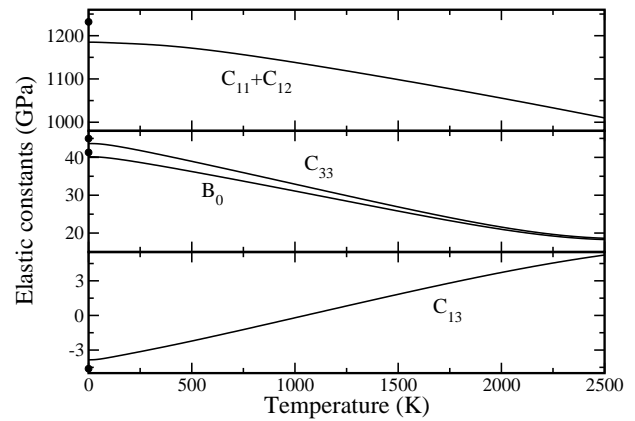}
\caption{\label{fig21}Elastic constants of graphite
($C_{11}+C_{12}$, $C_{13}$, $C_{33}$) and bulk modulus ($B_{0}$) as a function of temperature.
The filled circles (at 0 K) indicate 
their ground state values (as in Table~\ref{tab2}) before accounting for zero-point motion.}
\end{figure}

Results are shown in Figs.~\ref{fig20} and \ref{fig21} (diamond and graphite 
respectively). Again, the zero-point motion has a significant impact
on the elastic constants; the agreement with experimental data for the temperature 
dependence of the ratio of the bulk modulus of diamond to its 298 K value is excellent
(upper panel of Fig.~\ref{fig20}).

We note that the temperature dependence of the bulk modulus of diamond has already 
been obtained by Karch \textit{et al}\cite{Karch} using LDA calculations.

As final thermodynamic quantities, we present results on the heat
capacities for all the systems considered, at constant volume ($C_v$) and constant pressure ($C_p$).
$C_v$ has been computed using Eq.~\ref{heat}, in which we used at 
each temperature T the interpolated phonon frequencies calculated at 
the lattice constant(s) that minimize the respective free energy.
To calculate $C_p$, we added to $C_v$ the additional term 
$C_p-C_v= T V_0  B_0 \alpha^2_V$ where $V_0$ is the unit cell volume, 
$\alpha_V$ the volumetric thermal expansion and $B_0$ the bulk modulus. 
All these quantities were taken from our ab-initio results and evaluated 
at each of the temperatures considered. The difference between $C_p$ and $C_v$ 
is very small, at most about 2\% of the value of $C_v$ for graphite and 5\% for diamond.
Note that $C_p$ and $C_v$ shown on the figures are normalized
by dividing by the unit cell mass.

\begin{figure}
\includegraphics*[scale=0.32]{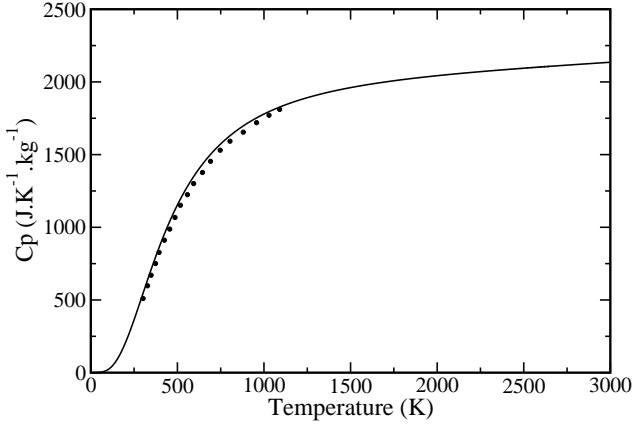}
\caption{\label{fig22}Constant pressure heat capacity for diamond (solid line).
Experimental results are from
Refs.~\onlinecite{Madelung} and \onlinecite{Victor} (circles), as reported by
Ref.~\onlinecite{Herrero-Ramirez}.}
\end{figure}
\begin{figure}
\includegraphics*[scale=0.32]{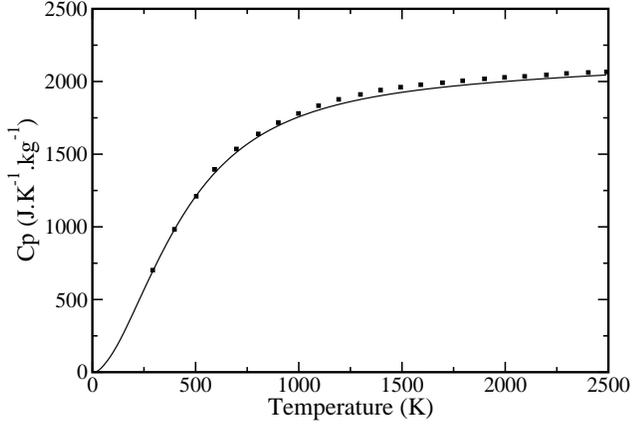}
\caption{\label{fig23}Constant pressure heat capacity for graphite 
(solid line). Experimental results are from
Ref.~\onlinecite{Hultgren} (squares), as reported by
Ref.~\onlinecite{Fried-Howard}.}
\end{figure}
\begin{figure}
\includegraphics*[scale=0.32]{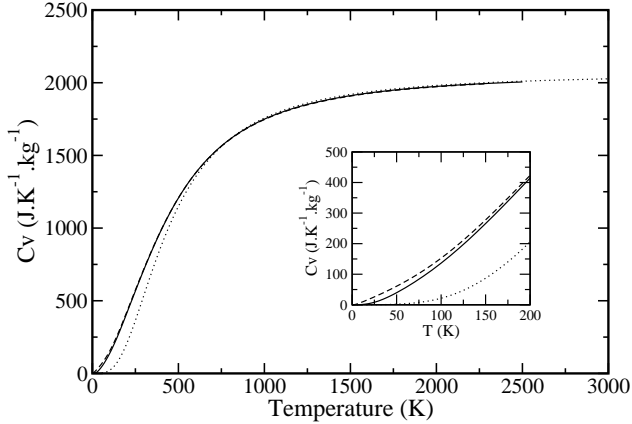}
\caption{\label{fig24}Constant volume heat capacity for graphite (solid line), graphene (dashed line)
and diamond (dotted line).
The inset shows an enlargement of the low temperature region.}
\end{figure}
The heat capacity of diamond, graphite and graphene are almost identical except at
very low temperatures. Agreement with experimental data is very good.

\section{\label{sec5}Conclusion}

We have presented a full ab-initio study of the structural,
vibrational and thermodynamical properties of diamond, graphite and
graphene, at the DFT-GGA level and using the quasi-harmonic approximation to 
derive thermodynamic quantities. All our results are in very good agreement 
with experimental data: the phonon dispersions are 
well-reproduced, as well as most of the elastic constants.
In graphite, the C$_{33}$ elastic constant
and the $\Gamma$ to $A$ phonon dispersions (calculated here with GGA for the 
first time) were found to be in good agreement with experimental results provided
the calculations were performed at the experimental $c/a$. Only the C$_{13}$
constant remains in poor agreement with experimental data.

The decay of the long-ranged interatomic force constants was
analyzed in detail. It was shown that
interactions in the (110) direction in diamond are longer-ranged
than these in other directions, as is characteristic of the
zincblende and diamond structures. For graphene and graphite, in-plane
interactions are even longer-ranged and phonon frequencies
sensitive to the truncation of the interatomic force constants.

Thermodynamical properties such as the thermal expansion,
temperature dependence of elastic moduli and specific heat were calculated
in the quasi-harmonic approximation. These quantities were all found to be in close agreement
 with experiments, except for the out-of-plane thermal expansion of graphite 
at temperatures higher than 150 K. 
Graphite shows a distinctive in-plane negative thermal-expansion coefficient 
that reaches the minimum
around room temperature, in very good agreement with experiments. This effect
is found to be three times as large in graphene. In both cases, the mode Gr\"uneisen 
parameters show that the ZA ``bending''
acoustic modes are responsible for the contraction, in a direct manifestation of the membrane
effect predicted by Lifshitz\cite{Lifshitz} in 1952. 
These distinctive features will likely affect the thermodynamical properties of 
single-walled and multiwall carbon nanotubes\cite{Schelling-Keblinski,Tomanek,Mounet-Marzari}.

\begin{acknowledgments}
The authors gratefully acknowledge support from NSF-NIRT DMR-0304019 and 
the Interconnect Focus Center MARCO-DARPA 2003-IT-674.
 Nicolas Mounet personally thanks the Ecole Polytechnique of Palaiseau (France) and 
 the Fondation de l'Ecole Polytechnique for their help and support.
\end{acknowledgments}

\end{document}